# VIBRATIONAL ENTROPY, CONFIGURATIONAL ENTROPY, AND THE FRAGILITY OF GLASSFORMING LIQUIDS.

L.-M. Martinez and C. A. Angell,
Dept. of Chemistry and Biochemistry,
Arizona State University,
Tempe AZ 85287-1604

**Understanding of fluidity and diffusion in viscous liquids is in a state of flux. Formerly considered as a problem in energy barrier crossing hence strictly kinetic in nature, it is now being suggested by molecular dynamics studies that the liquid diffusion process is a process dominated by thermodynamic factors. Related to this development, there exist two separate conflicts involving experimental data, and the resolution of these conflicts should do much to advance our understanding of the "glassy dynamics" problem. The first conflict concerns the relation between kinetic fragility (which relates to the peculiar temperature dependence of transport in viscous liquids) and thermodynamic fragility. Thermodynamic fragility has recently (Ito et al ref.1) been defined in terms of the temperature dependence of the excess entropy of liquid over crystal, scaled by the excess entropy at the glass transition temperature, and it has been correlated with kinetic fragility. This correlation has been vigorously contested [2] using dielectric relaxation data on some molecular liquids. We put the conflict in perspective using an extensive data set covering all classes of liquids while retaining the cases of ref 2. The *excess* entropy of the liquid, used in the thermodynamic quantity being correlated, has both configurational and vibrational components and this is the source of the second conflict. This concerns the relative importance of these two components. Experimental tests supporting the Adam-Gibbs "entropy" theory of the temperature dependence of transport in viscous liquids have mostly been made using the excess entropy, but recent simulation studies on the hard sphere fluid [3] SPC-E water [4] and mixed LJ [5,6] have shown, as in the A-G theory, a correlation with the configurational component. We show why, in laboratory studies, the correlation should be with both, inextricably. In the course of this demonstration we identify a key role for excess vibrational entropy in causing fragile behavior in ambient pressure liquids.**

Nearly two years ago Sastry et al [7] presented an "inherent structures" analysis of MD data on a simple binary liquid mixture ("mixed LJ") that does not crystallize in computer supercooling experiments. They demonstrated that the features of the liquid diffusivity and structural relaxation that have provoked most interest among theorists could be related to the static structure through the potential energy hypersurface features that the liquid explores with highest probability at different temperatures. The properties that were related to the energy surface features were the onset of "super-Arrhenius" behavior on cooling into the low diffusivity regime, and the critical temperature $T_c$ of the mode coupling theory of glass formation (obtained by power law fitting of the D vs T relation). An implication of this study was that the rate (per unit temperature) of excitation of the liquid, from low energies towards the "top of its energy landscape," would provide a measure of its fragility that was basically thermodynamic. This was quantified by Speedy who derived [8] an expression for thermodynamic fragility in which the excess entropy frozen in at the kinetic glass transition plays an important role. At the same time it appears to connect directly to the super-Arrhenius behavior that defines the kinetic fragility of the liquid.

The thermodynamic connection was also followed up by Ito et al [1] who suggested that the entropy data used by Kauzmann in his formulation of the paradox bearing his name, could be utilized to quantify the thermodynamic fragility. They suggested recasting the entropy data in a diagrammatic form that has the same character as the $T_g$-scaled Arrhenius plot commonly used to display the different degrees of kinetic fragility of liquids. This required the same scaling by the excess entropy frozen in at $T_g$ as occurs in Speedy's expression [8]. Ito et al observed, qualitatively, that the ordering of liquid fragility was the same in kinetic and thermodynamic manifestations.

This notion was immediately contested by Ngai and colleagues [2,9]. These authors suggested, from an analysis of the dielectric relaxation times and entropies of 9 molecular liquids [2], that the correlation was unreliable at best and, in the case of polymeric liquids [9], was without any merit at all. Their conclusion however clashes with the repeated analyses of the calorimetric and kinetic behavior of liquids (not polymers) of different classes that extract "ground state temperatures" from the data and find them to have very similar values. We refer to the Kauzmann temperature $T_K$ (where the supercooling liquid entropy extrtapolates to the crystal entropy) and the Vogel temperature $T_o$ (where the viscosity extrapolates to infinity by the Vogel-Fulcher-Tammann equation), (see, e.g., ref. 10 for some 30 cases where $T_o = T_K$ within +/- 3% despite several prominent exceptions). Since the ratio $T_g/T_o$ is often taken as a measure of fragility [11,12] it is clear that $T_K = T_o$ implies that kinetic fragility and thermodynamic fragility $T_g/T_K$ are the same, with some "glaring exceptions". This means that either the physically constrained analysis leading to $T_o \sim T_K$ is flawed or that the analysis of Ngai and co-workers, being limited to dielectric relaxation of molecular liquids, and containing some misplots, has distorted the overall picture.

If kinetic fragility is indeed dominated by thermodynamics in the "normal" case, then a most significant simplification of the whole fragility phenomenon (and concomitantly the topology of the potential energy surface) is at hand, so the problem demands some further analysis. Such an analysis is the purpose of the first and main part of this letter.

We start with an important proviso concerning the thermodynamic fragility. In the definition of this quantity in ref. 1, an important advantage of the kinetic fragility is missing. This is the manner in which the kinetic fragility is made independent of the properties of the crystal phase. By using $T_g$ as a reference temperature instead of the melting point, the kinetic fragility is made an entirely liquid state quantity. In the thermodynamic fragility analog defined by Ito et al. [1], this reference temperature scaling advantage is maintained. Unfortunately, however, in our attempt to select the important part of the entropy for liquid state behavior (i.e., the part in excess of the entropy of the fixed structure, which is usually taken as that of the crystal) we find ourselves still at the mercy of idiosyncrasies of the crystal phase. This is because changes of vibrational entropy on fusion will affect the entropy of fusion, to which the excess entropy of the liquid is always referred (*except* in computer simulations [3-5,13,14]).

This has been pointed out before, by Richet (15,16), in his consideration of the thermodynamics of geosilicate glassformers. Richet noted that it is not possible to find a systematic relation between the entropy of fusion and the liquid properties in geochemical melts because of the complexities of the crystal thermodynamics. The importance of Richet's observation has been brought home to the present authors by the finding that the archetypal strong liquid, $SiO_2$ itself, appears intermediate by its $S_g/S(T)$ behavior, though by the change of heat capacity at $T_g$, it is at the strong liquid limit as expected. This anomaly is a consequence of the fact that in silica, uniquely [15], the vibrational entropy of the glass is much less than that of the crystalline polymorphs. This has the consequence that the entropy of fusion, and therefore the computed excess entropy of the glass at $T_g$, (which scales the liquid entropy [1,7]) is confusingly small. This would not happen if the glass entropy could be assessed by either of the all-liquid routes used in computer simulation studies[3-5, 13, 14]. Comparable anomalies, having different origins, attend some other complex silicates. However, Richet showed that, if attention is focussed on the increase in heat capacity at $T_g$ (which, like the viscosity, is a purely liquid state property), then kinetic fragility and thermodynamic fragility judged by $C_{p(liquid)}/C_{p(glass)}$ at $T_g$ are systematically related for these mineral liquids - which are mostly "strong" in character.

A similar problem with chain polymer crystals, in which the force constant distribution must be even more heterogeneous than in silicate crystals, is perhaps the reason that attempts to correlate Vogel-Fulcher and Kauzmann temperatures in polymers have never met with much success. It might also explain why, for polymers, Roland et al. [9] found such disarray in the correlation we are discussing in this letter. Until we can obtain the entropy of the glass at $T_g$ using the ideal gas reference state [3-5, 13,14], we must expect to find some serious anomalies. We do not think we should let this prevent us from making an attempt to see the broader picture. It is hoped that the conclusions we are lead to in this letter will justify our tolerance of some cases that, without the proviso just made, might seem to sustain the conclusions of refs. [2,9]. Why not avoid the crystal problem by reverting to the use of $\Delta C_{p,liquid}/\Delta C_{p,glass}$, a purely liquid state property, as in the past? Speedy's argument [8], and the Adam-Gibbs equation [17] (the form of which has now been strongly supported by MD simulations [3-6]) show that the scaled quantity is required.

With the foregoing proviso in mind, we will now show that comparison of viscosity data (which unlike dielectric data are available on liquids of all types) with thermodynamic data on the same liquids, broadly confirm the original suggestion, as expected from the earlier $T_o/T_K \sim 1.0$ finding. It is important to note that $T_o = T_K$ is also a finding of MD simulations on well-defined systems, using all-amorphous phase calculations [13,14,3-6]). We assess data on some 25 liquids from the fields of geochemistry (liquid silicates), covalent semiconductors (liquid chalcogenides), molecular liquids and molten ionic hydrates, salts and oxides, including all the systems analyzed by Ngai and Yawamuro [2]. We find a wide-ranging correlation of kinetic fragility (obtained from viscosity data) with thermodynamic fragility data (obtained from *excess* entropy data). While a simple 1:1 correlation is not found and indeed is not expected [17,18], the broad correlation we find has important implications. We address these briefly in the second part of this letter which focuses on the fact that our correlation of kinetic fragility is with a thermodynamic fragility defined from the *excess* entropy - which is only partly configurational in content. (The excess entropy contains other components in proportions that were originally examined by Goldstein [19,20], and that have recently excited the interest of several groups, [21-23]).

Figs. 1 and 2 show the kinetic fragility and thermodynamic fragility plots of the liquids that are listed in the legends (in order of increasing fragility). All temperatures are scaled by the temperatures of the calorimetric onset glass transition for the substance, measured during upscan at 10K/min (after prior continuous cooling at the same rate). This corresponds to the temperature at which the *enthalpy* relaxation time is approximately 200 s [11].

It is to be noted that the order of the curves is the same in each plot with only a few exceptions. Unlike ref. 2, the present data set covers the whole known fragility range. It also incorporates members of every class of glassforming system except metals (for which appropriate compound entropy data are unfortunately unavailable). The case of ethanol, as included in Kauzmann's original plot, is actually a solid (rotator phase) hence its viscosity cannot be measured at zero shear stress and no data for this phase can be included in Fig. 1. We note in Fig. 2 that this rotator phase is less fragile than the other molecular liquids. A separate study of rotator phases, using reorientation times to determine the kinetic fragilities, should be profitable (since a wide range of deviations from the Arrhenius Law are known in these cases).

To make a more quantitative comparison, we may use the intersections of the individual plots with the mid-range line indicated for each plot. This gives us the $F_{1/2}$ fragility, the advantages and measurement of which have been discussed elsewehere [12,22] for the simpler case of relaxation times. Relaxation time fragility plots are simpler because they all pass through the same point at the reference temperature. The viscosities of fragile molecular liquids tend to be less than the canonical $10^{12}$ Pa.s at $T_g$. However this metric is the best that can be applied to viscosity data while retaining the calorimetric $T_g$ as scaling temperature. Elsewhere [24] we include consideration of the mechanical relaxation times that are available for a smaller set of liquids. (Viscosities are proportional to shear relaxation times via a Maxwell relation in which the proportionality constant [G, the high frequency shear modulus] has only a weak temperature dependence). For the thermodynamic fragility it is preferable to use the 3/4 line, since this avoids the need for extrapolations in the case of strong liquids.

The thermodynamic and kinetic fragilities, defined in this way, are compared in Fig. 3. While the "glaring exceptions" are of course still present, the majority of the substances analyzed show thermodynamic fragilities that are correlated with their kinetic fragilities. The exceptions identified in ref. 2 weighs less heavily amongst the larger number of simply related cases. Note the wide variety of different liquid types of intermediate fragility that cluster around selenium in Fig. 3.

Figure 3 demonstrates a much closer correlation than we would have expected from the application of the Adam-Gibbs equation [17] which contains a specifically kinetic parameter (see below). We previously believed [18] that this parameter would exert an independent control of the fragility but it now seems that the effect is minor. However before anything can be made of this simplification in the phenomenology we must face the fact that our correlation is not the one that even a simplified Adam-Gibbs equation would predict. This is because the excess entropy [25] we have used to construct Fig. 2 is not the quantity that should appear in the Adam-Gibbs equation, notwithstanding the fact that it is the quantity that has been used in most of the experimental tests of the Adam-Gibbs equation. The second part of this letter analyzes the important implications of this difference.

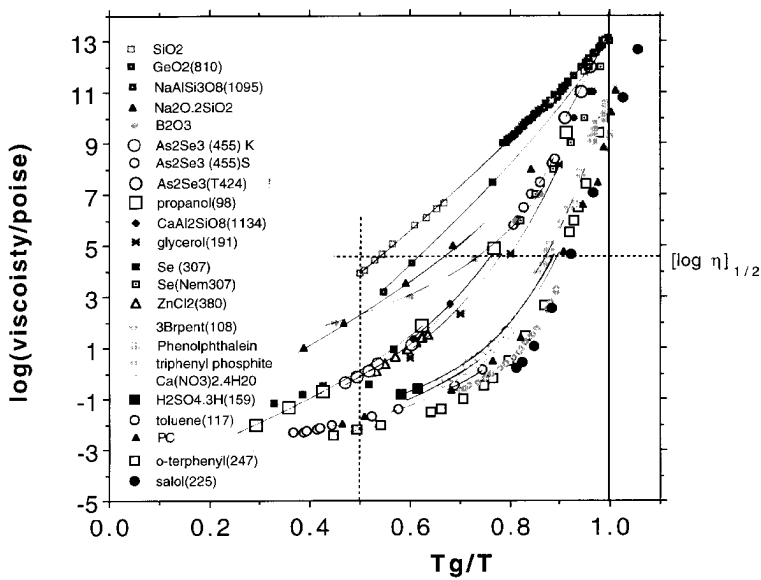

**Figure 1.** $T_g$-scaled Arrhenius plot for liquids of different classes for which thermodynamic data are also available. The scaling parameter is the normal calorimetric $T_g$ at which the enthalpy relaxation time is ~200 s. The horizontal line are drawn at 1/2, of the way between $10^{13}$ poise characteristic of the glass transition for non-fragile liquids and $10^{-4}$ poise (which is the roughly common high temperature limiting value). The log $\eta_{1/2}$ line is used to obtain the $F_{1/2}$ fragility (kinetic) by the definition $F_{1/2} = 2T_g/T_{1/2} - 1$ which is recommended as the most reliable metric of this deviation from Arrhenius behavior, and is at the core of the fragility concept. Note that the kinetic fragilities of liquids whose viscosities are less than $10^{12}$ Pa.s at their calorimetric $T_g$ (due to small shear moduli or viscosity/structure decoupling) are exaggerated by this construction.

The Adam-Gibbs equation for the relaxation time of a dense liquid [17], which has been tested correctly and found to be of valid form in three recent computer simulation studies [3-5], has the form

$$\tau (\sim TD^{-1}) = \tau_o \exp(C/TS_c) \qquad (1)$$

where D is the diffusivity (and the T multiplier takes into account the Einstein relation between diffusivity and mobility), C is a constant which contains an energy barrier per particle, and $S_c$ is the configurational entropy. $S_c$ is that part of the entropy of a pure liquid which is determined by the plethora of possible distinct packing states accessible at the temperature T. It is related logarithmically to the number of minima on the Goldstein-

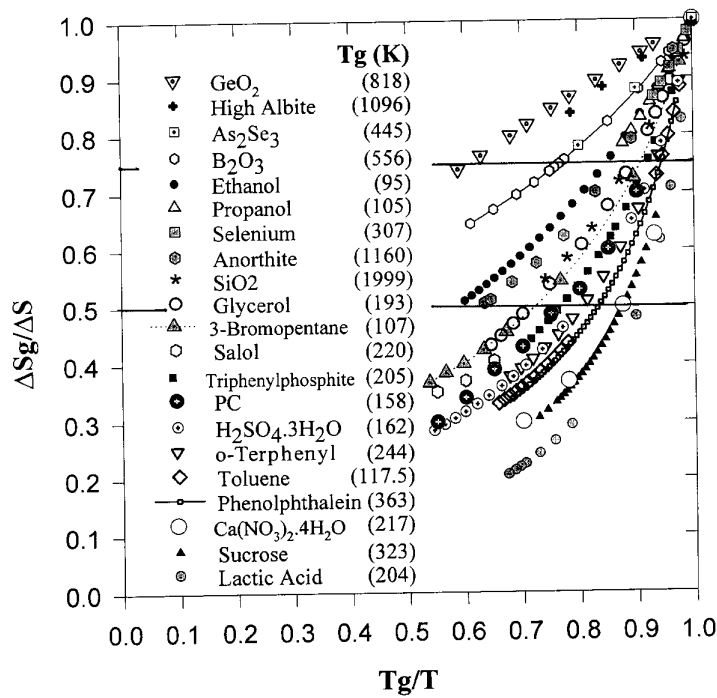

**Figure 2.** Excess entropy vs. $T_g$-scaled temperature plots. The entropy in excess of the crystal $\Delta S$ at different temperatures $T_g/T$ above $T_g$ is shown scaled by the excess entropy at $T_g$, $\Delta S_g$, so that the figure has a form similar to that of Figure 1. The lines drawn at the 0.5 and 0.75 marks are used to obtain thermodynamic fragilities $F_{1/2}$ and $F_{3/4}$ for comparison with the kinetic quantities. The latter is preferable, as it does not require any data extrapolations (for strong liquids) for its determination.

Stillinger potential energy hypersurface of the system's configuration space [26,27]. It is accessible by computer simulation studies when correctly performed [3-5]. While it has often been assumed that $S_c$ can be approximated by the difference in entropy between liquid and crystal, it has been known since Goldstein's 1976 analysis [19, extended in 20] of entropy change in glasses between $T_g$ and 0K, that this is a rather poor approximation in many cases. This has been emphasized by the recent extension of this analysis by Johari [23]. According to these studies the configurational component of the excess entropy may be as small as 40% of the excess entropy, with no pattern obvious for different types of liquids. A reason for ignoring these observations has no doubt been the fact that the Adam-Gibbs equation usually tests well (i.e linearizes the relaxation time or diffusivity data near $T_g$) using the easily accessible *excess* entropy [28-33,10] . Fig. 3 provides another example of this success. Yet MD simulation studies of both SPC-e water [4] and mixed LJ liquid [5] find that Eq. (1) holds as written. In neither case is the excess entropy equal to the configurational entropy. How is it that the Adam-Gibbs equation can be satisfied for both quantities?

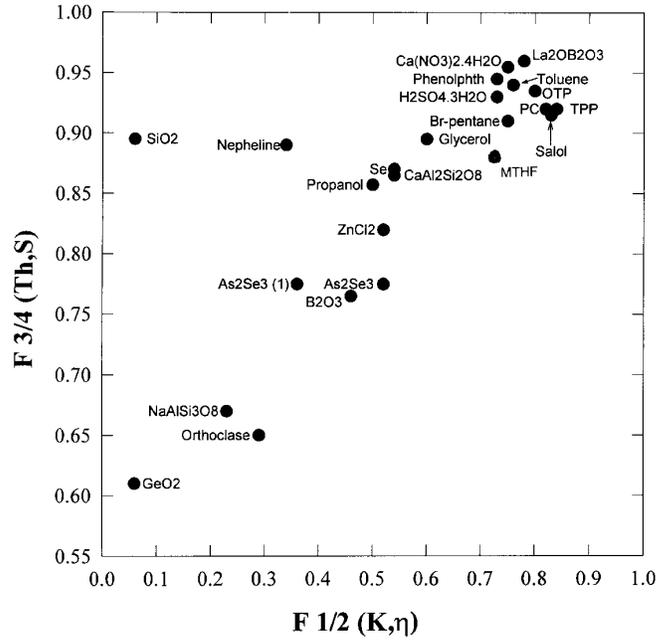

**Figure 3**. Correlation of the thermodynamic and kinetic $F_{1/2}$ fragilities from Figs 1 and 2, sucrose and lactic acid excepted. The thermodynamic fragilities of strong liquids, e.g. $SiO_2$, are very susceptible to distortion due to vibrational entropy differences of crystal and glass, and would be better assessed using heat capacity jumps at the glass temperature (see text). This figure contains additional data not included in Figs 1 and 2 for reason of crowding. The thermodynamic data appear already in Fig. 3 of ref. 2. The present figure contains all of the examples included in ref. 2, Fig. 3 insert, which are concentrated in the top right-hand corner of the present figure (glycerol and above). We have added inorganic liquids to this area. In ref 2, Fig. 3 inset, MTHF is the outstanding exception, Br-pentane was misplotted, and only one of the two toluene entries is relevant. In ref. 10, propanol, salol and phenolphthalein were exceptions. On the scale of the present figure these deviations are less prominent.

The important answer to this question must of course be that, as the temperature rises above $T_g$, both $S_c$ and $S_{ex}$ change at the same rate. When the rate of change is high, the liquid is fragile (Fig. 2). From these two statements we can deduce a lot about fragility, We show in detail elsewhere [24] how an increase in vibrational entropy in excess of the crystal as the structure changes above $T_g$ must promote a corresponding increase in configurational entropy, at least in the range over which the experimental tests of the Eq. (1) [25 -30] have been made (between $T_g$ and the melting point at about ~1.5$T_g$). For the purposes of this letter we will just present the argument graphically, in Fig. 4, using the concept of energy landscapes as developed in refs 3,26,27 and 34. In brief, the generation of extra vibrational entropy in the liquid due to the change in shape of the inherent structure basins (depicted in Fig. 4(a)) as the system "visits" higher energy basins at higher temperatures , provides an additional drive to higher energies. This is dictated by the Gibbs free energy function G=H-TS which must be minimized at each temperature. Note that, because we are discussing the results of Figs.1-3, we are depicting the energy profile for a constant pressure system which therefore has one extra dimension, volume, relative to the usual isochoric landscape (hence we refer to Gibbs, rather than Helmholtz, free energy). The more rapidly the landscape is ascended the more rapidly the configurational entropy is excited, Fig. 4(b), hence the more fragile the liquid must appear according to Eq. (1). It is the change in basin shape with increasing temperature that must therefore be a key ingredient in determining the fragility in liquids studied at constant pressure. It will act in concert with the klnW  (W = "number of glasses"[8]) factor identified by Speedy [8]  and by Sastry [6].

For studies conducted at constant volume, as is customary in computer simulations [3-6], a very different scenario is revealed by Sastry's recent study [6]. There the excess vibrational entropy can play an opposing role, but a quantitative relation between thermodynamic fragility (based on the T-dependence of the configurational entropy) and the kinetic fragility (based on the T-dependence of the diffusivity), is confirmed. This important difference needs to be borne in mind when comparing results of simulations and experiments.

Fig. 4(b) shows how extrapolations of either total excess entropy, or the configurational component of the excess entropy, should yield the same Kauzmann temperature.  Equally well, it shows why (i) Eq. (1) should represent the data, irrespective of whether excess entropy or configurational entropy is used and why (ii) in correlations of thermodynamic fragility with the kinetic quantity, excess entropy should serve as well as configurational entropy.

Finally it should be noted that Goldstein, in his paper identifying the non-configurational components of the excess entropy at the glass transition, suggested anharmonicity as a more likely source of the non-configurational excess heat capacity than low frequency vibrations. We could recast our above arguments in terms of anharmonic contributions to the entropy though it involves an additional level of complexity in the basin shape problem. In any case, the answer may not be simply one *or* the other because low frequency modes are invariably the more anharmonic. The two possibilities are closely related, and probably very much entangled. This leaves, as a key question for future work, the identification of  features of the pair potential for a liquid that  lead to excess vibrational heat capacity - or anharmonicity - in the liquid over crystal, and thereby to fragile liquid behavior. It  is a question that computer simulation studies hopefully will be able to answer.

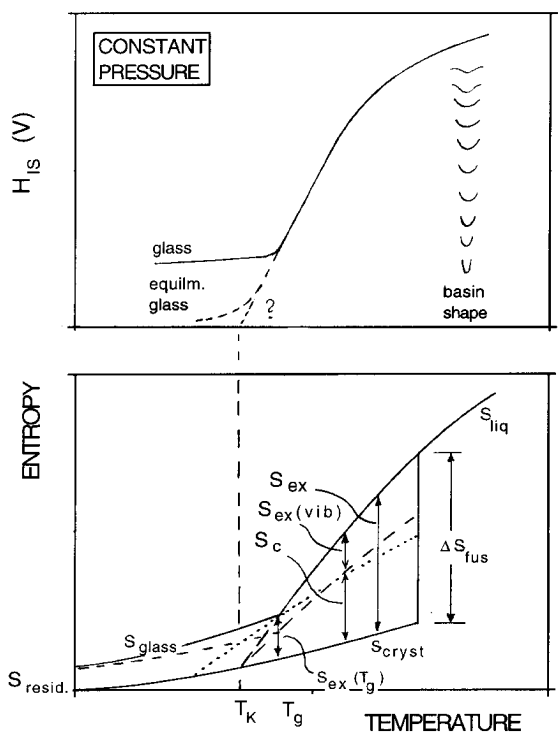

**Figure 4**. (a) Depiction of the inherent structure enthalpy profile for a one component system during heating from 0 K to above the melting point *at constant pressure*. Changes in shape of the basins visited with greatest probability at each temperature and volume are depicted to the right of the excitation profile. These become progressively shallower as temperature increases and the system expands.
(b) Entropy vs temperature for the system depicted in part (a), contrasting the monotonic increase of the system, when in the crystal basin, with the behavior of the amorphous phase. The glass is depicted as having excess entropy $S_{residual}$ at 0 K but a larger $S_{ex}$ at $T_g$. This is due to the vibrational density of states of the glass being richer in low frequency modes (particulalry near the boson peak) than that of the crystal. Above $T_g$ the vibrational entropy increases more rapidly than below $T_g$ due to the system access to basins of different shape as depicted in part (a). The access to additional vibrational entropy provides an additional T S drive which must be balanced by more rapid increase in enthalpy, than in the case where all basins have the same shape. The division of the excess entropy into its configurational and vibrational components is made by the heavy dashed line. In the case of constant basin shape, the drive to higher enthalpies comes only from the configurational entropy, viz., $TS_c$. The change in $S_{ex}$, which is also $S_c$ in this case, is depicted as a dotted line. Note that, for variable basin shape, the increased rate of excitation of vibrational entropy has increased the rate of generation of *configurational* entropy [30], as well as that of the excess entropy. $T_g/T_K$ must therefore be smaller for this case, hence the fragility larger.

## Acknowledgements

This work was supported by the NSF DMR Solid State Chemistry program.